\documentclass[11pt,a4paper]{article}
\usepackage{jheppub}
\usepackage{mathtools,amsmath,amsfonts}
\usepackage{cleveref}
\usepackage[T1]{fontenc}
\usepackage[latin9]{inputenc}
\usepackage[sort&compress]{natbib} 
\usepackage{silence}
\usepackage{babel}
\newcommand{\overleftrightsmallarrow}{\mathpalette{\overarrowsmall@\leftrightarrowfill@}}
\usepackage{caption} 
\usepackage{subcaption}
\usepackage{graphicx} 
\usepackage[toc,page]{appendix}

\title{Baby skyrmion in two-component holographic superfluids}

\author[a]{Shunhui Yao,}
\author[a,b]{Yu Tian}
\author[a]{Peng Yang}
\author[c]{Hongbao Zhang}
\affiliation[a]{School of Physical Sciences, University of Chinese Academy of Sciences, 
\\ Beijing 100049, China } 
\affiliation[b]{Institute of Theoretical Physics, Chinese Academy of Sciences,\\
Beijing 100190, China}
 \affiliation[c]{Department of Physics, Beijing Normal University, \\
Beijing 100875, China}
 \emailAdd{yaoshunhui15@mails.ucas.ac.cn}
 \emailAdd{Corresponding author: ytian@ucas.ac.cn}
 \emailAdd{Corresponding author: yangpeng18@mails.ucas.ac.cn}
 \emailAdd{Corresponding author: hongbaozhang@bnu.edu.cn}

\abstract{In the two-component Ginzburg-Landau theory of superfluidity, a pair of fractional vortices form a composite type of topological defect, usually referred to as a baby skyrmion. In this paper, we initiate the construction of such a baby skyrmion in the holographic model of two-component superfluids. As a result, two types of baby skyrmion  configurations are found, where the monopole-type of one is constructed directly by solving the static equations of motion while the dipole-type of one is obtained by resorting to the time evolution method. In addition, we find that the existence of these two types of baby skyrmion depends on the inter-component coupling, reminiscent of the situation in the baby skyrmion  model. 
}

\begin{document}
\maketitle

\section{Introduction}

Topological defects, which occur ubiquitously in spontaneous symmetry breaking, have been discussed in various contexts ranging from cosmic strings to superconductivity, where Landau's paradigm of order parameters is commonly adopted to study the mechanism of phase transition. In the 2D Bose-Einstein condensates described by the Ginzburg-Landau (GL) theory, quantized vortices are known as elementary topological defects that appear commonly in both superconductor and superfluid systems. In superconductivity theories, the quantum vortex is called Abrikosov vortex\cite{abrikosov1957magnetic}, which carries, according to London's Law, a quantized magnetic flux, while in the context of superfluids, the counterpart corresponds to Onsager-Feynman quantization of superfluid velocity\cite{feynman1955chapter,onsager1949statistical}. Both cases of quantization could be attributed to the single-valuedness of the order parameters, and thus could be classified by the topological charge, called winding number. 

To make a straightforward generalization of the ordinary GL theory, one can extend the one-order-parameter model to the multi-component cases, where fractional vortices are instead considered as the fundamental type of topological defects, featuring the violation of the ordinary flux-quantization or Onsager-Feynman quantization in one-order-parameter GL theories\cite{Babaev:2001hv,babaev2007violation,2020PhRvA.101a3630R,2021PhRvA.103b3311R}. In particular, in the two-component case,  such topological defects can be understood as baby skyrmions  by constructing a mapping from one $S^2$ (the 2D spatial plane with infinity included) to the other $S^2$ comprised of the 3D unit vectors with its three components given by the expectation values of Pauli matrices in the two-component condensates\cite{Garaud:2013vya}. Accordingly, the resulting baby skyrmions can be characterized by the topological charge associated with the second homotopy group of the sphere, called  skyrmion  number\cite{Piette:1994mh}. 
On the other hand, via the AdS/CFT correspondence\cite{Maldacena:1997re,Gubser:1998bc,Witten:1998qj}, the Abelian-Higgs model of holographic superfluids and superconductors was introduced in \cite{Hartnoll:2008vx,Hartnoll:2008kx,Herzog:2009he},  with the bulk hairy black hole dual to the the phase of spontaneously broken $U(1)$ symmetry on the boundary\cite{Gubser:2008px}. Since then, such a holographic model  has been extensively studied.  In particular, not only was the static configuration of a single vortex constructed with its characteristic features identified\cite{Keranen:2009re,Montull:2009fe,Domenech:2010nf,Montull:2011im,Salvio:2012at,Montull:2012fy,Dias:2013bwa,Salvio:2013jia,Lan:2017}, but also the real time holographic vortex dynamics was  explored\cite{Adams:2012pj,Du:2014,Lan:2016,Wittmer:2020mnm,Yan:2022jfc}. 

Just like the GL theory, the above
holographic model of superfluids/superconductors can also be extended to the  two-component cases\cite{He,Du:2015zcb}, where the vortex-like topological defects were revealed and the real time dynamics of vortices were investigated\cite{Wu:2015sqk,Yang:2019ibe}.  A natural question along this line is whether there exist baby skyrmions  in such two-component holographic superfluids/superconductors. The purpose of this paper is intended to address this issue by focusing on the topological charge $\mathcal{Q}=1$ baby skyrmion. As a result, not only do we succeed in constructing the dipole-type of baby skyrmion, but also the monopole-type of baby skyrmion. Furthermore, we find that the existence of such baby skyrmions depends on the bulk inter-component coupling parameter $\nu$, where the monopole-type of baby skyrmion  can only exist in the region with $0<\nu<\nu_c$ and the dipole-type of baby skyrmion  can only exist in the region in which $\nu$ is negative, which 
is reminiscent of the result for the baby skyrmion  model, where the existence of the analogous baby skyrmions also depends on the specific interaction\cite{Leask:2021hzm}.

This paper is organized as follows. In the subsequent section, we shall introduce the holographic model of two component superfluids we adopt in this paper and elaborate on the two types of baby skyrmion   under our consideration. In Section \ref{sec1}, we will construct the monopole-type baby skyrmion   by solving the corresponding static equations of bulk fields. While in Section \ref{section2}, we shall resort to the time evolution method to search for the static configuration of the dipole-type baby skyrmion   via dynamic evolution. We conclude our paper with some discussions. 

\section{Baby skyrmion   configurations in Holographic superfluids}

\subsection{Holographic two-component superfluid model}

We first introduce the holographic model of two-component superfluids. As such, we consider an Abelian-Higgs model in (3+1)D asymptotic anti-de-Sitter spacetime with $L$ the AdS radius\cite{Yang:2019ibe,Wu:2015sqk}, i.e.,
\begin{equation}
	S=\int_{M} \sqrt{-g} d^{4} x\left[\frac{1}{2 \kappa^{2}}\left(R+\frac{6}{L^{2}}\right)-\frac{1}{e^{2}}\left(\frac{1}{4} F_{\mu\nu}F^{\mu\nu}+ (D^\mu\Psi)^\dagger D_\mu \Psi+ V(\Psi)\right)\right] \label{eq:1}.
\end{equation}
Here $\Psi=(\Psi_1,\Psi_2)$ is a scalar doublet, $F_{\mu\nu}=2\partial_{[\mu}A_{\nu]}$ is the field strength of the $U(1)$ gauge field $A$, and $D_\mu=\partial_\mu-iA_\mu$  is the $U(1)$ covariant derivative operator with $e$ and $\kappa=\sqrt{8\pi G_N}$ the coupling constants of gauge field and gravitational field respectively, where $G_N$ is the Newton constant\footnote{Here for simplicity, we have implicitly assumed that the couplings of $\Psi_1$ and $\Psi_2$ with the $U(1)$ gauge field are the same.}. In addition, 
we choose the potential as  $V(\Psi)= m_1^2 |\Psi_1|^2+ m_2^2|\Psi_2|^2+\nu|\Psi_1|^2|\Psi_2|^2$ with $m_1$ and $m_2$ the mass parameters of the scalar fields and $\nu$ the coupling parameter between the scalar doublet\footnote{ Unlike its Ginzburg-Landau counterpart  in \cite{Garaud:2013vya}, the bulk potential here does not include the inter-band Josephson coupling term $V_\textbf{Josephson}\propto\Psi_1^*\Psi_2+\Psi_2^*\Psi_1$, as the presence of this term is equivalent to a shift of mass parameters of the two scalar fields after a diagonalization of the quadratic part of Lagrangian. }.

The model will be treated under the probe limit. Namely, the back reaction of the matter field onto the background geometry is neglected, which can be achieved as the coupling strength of gravitational field is small enough compared with that of gauge field, i.e. $\kappa\to 0$ or $e\to \infty$. Accordingly, the matter sector can be living in a fixed background geometry. Such a  background metric for the model of holographic superfluids at finite temperature is given by the  Schwarzschild-AdS black-brane 
\begin{equation}
	d s^{2}=\frac{L^{2}}{z^{2}}\left(-f(z) d t^{2}+\frac{1}{f(z)} d z^{2}+d x^{2}+d y^{2}\right),
\end{equation}
where $f(z)=1-(z/z_h)^3$ with $z_h$ the black hole event horizon. The Hawking temperature of the black-brane is given by $T=3/(4\pi z_h)$, which is identified with the temperature of the dual superfluid system by holography. Without loss of generality, below we shall work with the unit in which $L=1$ and set $z_h=1$\footnote{Note that our theory is scale invariant, so only the dimensionless ratio quantities are meaningful. Our findings can be represented in terms of the corresponding dimensionless quantities simply by  multiplying the appropriate power of the temperature ($T=\frac{3}{4\pi}$). For example, $\mu\rightarrow \mu/T, r\rightarrow rT, t\rightarrow tT, \langle\mathcal{O}_i\rangle\rightarrow \langle\mathcal{O}_i\rangle/T^{\Delta_{+i}} $. But this reconstruction does not affect the main result we want to convey in this paper, so we simply present our result in terms of the dimensional quantities. }.

With the asymptotic expansion of the bulk scalar fields near the AdS boundary $z=0$ into two modes, namely the non-normalizable mode $\phi_{+}$ and the normalizable mode $\phi_{-}$,
\begin{equation}
\Psi_i=z^{\Delta_{-i}} \phi_{-i} +z^{\Delta_{+i}} \phi_{+i}+ \cdots,
\end{equation}
one can read off the expectation value of the dual boundary scalar operator $\langle\mathcal{O}_i\rangle=\phi_{+i}$, where $\Delta_{\pm i}=3/2\pm\sqrt{9/4+m_i^2}$ with $\Delta_{+i}$ the scaling dimension of the corresponding scalar operator $\mathcal{O}_i$. In what follows, we shall focus on the case with $m_i^2=-2$ for simplicity.  While $\phi_{-i}$ on the boundary is regarded as a classical source in the generating functional of the correlation function of the scalar operator $\mathcal{O}_i$. When we consider the vacuum condensation of the system, the source $\phi_{-i}$ is switched off. Similar to the ordinary GL theory,  the field $\phi_{+i}$ serves naturally as the order parameter of the holographic superfluid system. Likewise, we can perform an asymptotic expansion for the gauge field,
\begin{align}
  A_t&=\mu-\rho z+\mathcal{O}(z^2)\\
  A_i&=a_i +b_i z+\mathcal{O}(z^2) ,\quad i=x,y,
\end{align}
where we have chosen the axial gauge $A_z=0$. According to the holographic dictionary, $\mu$ and $\rho$ are interpreted as the chemical potential and charge density on the boundary, respectively. In addition, the spatial component $a_i$ is related to the superfluid velocity, and $b_i$ corresponds to the charge current density on the boundary\cite{Herzog:2009he}. 
As for the homogeneous and isotropic holographic superfluid without the inter-component coupling, it has been shown in \cite{Yang:2019ibe} that when $\mu>\mu_c=4.07$, the vacuum expectation value of $\mathcal{O}_i$ becomes nonzero, indicating that the system undergoes a spontaneous breaking of the global $U(1)$ symmetry to the superfluid state from the normal fluid state. For our purpose, below we shall work exclusively with $\mu=5.0$, which gives rise to the superfluid phase for the whole regime of the parameter space of the coupling $\nu$ we are considering. 


\subsection{Baby skyrmion  constructed out of vortices}

\paragraph{From vortex to fractional vortex}
As mentioned in the introduction section, there exists a new type of topological defect in the 2D two-component GL theories, the fractional vortex, which differs from the ordinary vortex in the sense that the norm of the scalar doublet is everywhere none-zero\cite{Garaud:2013vya}. In particular, the asymptotic behavior of the order parameter in the vicinity of the fractional vortex core goes like
\begin{align}
 \quad \psi \sim \binom{C\neq0}{r^{n}e^{in\theta}} \quad \text{or} \quad  \binom{r^{n}e^{in\theta}}{C\neq0}  \quad \text{when } r\to 0 ,
\end{align}
which is different from that in the vicinity of the ordinary vortex core
\begin{align}
 \psi \sim (re^{i\theta})^n\binom{C_1}{C_2},
	\quad |C_1|,|C_2| <\infty \quad \text{when } r\to 0 ,
\end{align}
where $(r,\theta)$ is the local polar coordinate with the vortex core located at $r=0$, and $n$ is an integer number. Accordingly, unlike the ordinary vortex, where the two components share the same winding number $n$ near the vortex core, the fractional vortex carries the winding number $n$ for one component but carries no winding number for the other one. This justifies why it is named the fractional vortex.



\paragraph{From fractional vortex to baby skyrmion  }
According to \cite{Babaev:2008zd,Babaev:2001zy,Garaud:2013vya}, the two-component GL theory can be reformulated into a nonlinear sigma model\cite{Faddeev:1975tz,faddeev1995einstein}, where the baby skyrmion model of a 3D unit vector field $\mathbf{n}$ is referred to as the pseudo-spin:
\begin{equation} \label{pauli}
	\mathbf{n}=\frac{\psi^\dagger\mathbf{\sigma} \psi}{\psi^\dagger\psi}, \quad \sigma^i ~\text{is the \textit{i}th Pauli matrix}.
\end{equation}

A baby skyrmion is defined as a topologically nontrivial configuration of the $\mathbf{n}$ field, which is characterized by the homotopy class of the mapping $\varphi $ from the compactified coordinate space $\mathbb{R}^2 \cup {\infty}\simeq S^2 $ to the target sphere $S^2$ : $  \mathbf{x}\mapsto \mathbf{n}$. The corresponding topological charge  
\begin{equation}
	\mathcal{Q}:=\frac{1}{4\pi}\int_{\mathbb{R}^2} \varphi^* \omega =\frac{1}{4 \pi} \int_{\mathbb{R}^{2}} \mathbf{n} \cdot \partial_{x} \mathbf{n} \times \partial_{y} \mathbf{n} ~\mathbf{d}^2\mathbf{x}, \label{eq:7}
\end{equation}
is called the skyrmion number , where $\omega$ is the standard volume form of $S^2$. The skyrmion number  can be an arbitrary integer, in accordance with the fact that the second homotopy group of sphere is given by $\pi_2(S^2)\simeq \mathbb{Z}$.

As shown in \cite{Babaev:2008zd,Babaev:2001zy,Garaud:2013vya},  a baby skyrmion   can be composed of two confined fractional vortices. Hence it is a composite type of topological defect in the two-component GL theory. However, besides the dipole-type baby skyrmion  discussed in \cite{Garaud:2013vya}, we argue that there is also a monopole-type baby skyrmion. 
To be more specific, for the $\mathcal{Q}=1$ dipole-type baby skyrmion, the order parameter has the profile
\begin{equation}
	\psi=\binom{f_1(\mathbf r)e^{i\theta_1}}{f_2(\mathbf{r})e^{i\theta_2}},\quad \theta_i=\arctan{\frac{y-y_i}{x-x_i}} ,\quad i=1,2,\label{eq:5}
\end{equation}
where $\mathbf{r}_i=(x_i,y_i)$ and $ f_i$ stand for the center location and the core function of the two fractional vortices such that $f_i(\mathbf{r})\to |\mathbf{r}-\mathbf{r}_i|$ as $\mathbf{r}\to\mathbf{r}_i$ while $f_i$ approaches  the magnitude of the homogeneous condensate at spatial infinity.
In contrast, for the $\mathcal{Q}=1$ monopole-type baby skyrmion, the order parameter has the profile
\begin{equation}
	\psi=\binom{f_1(\mathbf{r})}{f_2(\mathbf{r})e^{i\theta_2}}, \quad \theta_2=\arctan{\frac{y-y_2}{x-x_2}}
	\label{eq:2}
\end{equation}
with $f_2$ behaving as before but $f_1\to 0$ at the spatial infinity.

Both types of baby skyrmion   carry a topological charge $\mathcal{Q}=1$. Here we like to present a heuristic argument for it. According to Eq. (\ref{pauli}), we have
\begin{equation}
    \mathbf{n}=(\frac{\zeta+\bar\zeta}{1+\zeta\bar{\zeta}},-i\frac{\zeta-\bar\zeta}{1+\zeta\bar{\zeta}},\frac{1-\zeta\bar\zeta}{1+\zeta\bar{\zeta}})
  \label{eq:4}
\end{equation}
with $\zeta$ the ratio of the two components of the scalar doublet, namely,  $\zeta=\psi_2/\psi_1$. Whence we can express $\zeta$ in terms of $\mathbf{n}$ as follows
\begin{align}
\zeta&=\frac{\mathbf{n}^1+i \mathbf{n}^2}{1+\mathbf{n}^3},
\end{align}
which amounts to saying that the mapping $\varphi :\mathbf{x} \mapsto \mathbf{n}$ can be represented by the complex function $z \to \zeta(z)$ from the coordinate space to the target sphere  $S^2 \simeq \mathbb{CP}^1$ with $z=x+i y$ the complex coordinate on $\mathbb{R}^2$. Then the skyrmion number  $\mathcal{Q}$ can be written as
\begin{equation}
    \mathcal{Q}=\frac{1}{4\pi}\int \frac{ 2 i d \zeta d\bar\zeta }{(1+\zeta\bar\zeta)^2},
\end{equation}
which is identical to the degree of the mapping. For the case of monopole-type baby skyrmion   \eqref{eq:2}, the mapping is homotopic to the linear function  $\zeta= z-z_c$ with the vortex core $z_c$ and the infinity mapped to the north pole and south pole of the target sphere, respectively. While for the case of dipole-type baby skyrmion   \eqref{eq:5},  the mapping is homotopic to the function $\zeta=\frac{z-z_1}{z-z_2}$ with the two vortex cores $z_1$ and $z_2$ mapped to the north pole and the south pole of the target sphere, respectively. So both types of baby skyrmion have $1$ as the degree of the mapping. 

In the following sections, we are going to construct both types of baby skyrmion   in the holographic two-component superfluid model by numerics.

\section{Monopole-type baby skyrmion  }\label{sec1}

\subsection{Ansatz for static solutions and relevant boundary conditions} 
Motivated by the previous discussions, the ansatz for the bulk static scalar doublet is of the form: 
\begin{equation} 
	\Psi(z,r,\theta)=z\binom{u_1(z,r)}{u_2(z,r)e^{i\theta}},\label{eq:ans1}
\end{equation}
where the vortex is located at the origin.
For the static gauge field sector, the ansatz can also be made to respect the rotational symmetry as follows
\begin{equation}
 A_t=A_t(z,r),~ A_r=0,~ A_\theta=A_\theta(z,r).\label{eq:ans2}
\end{equation} 
Substituting Eq. \eqref{eq:ans1} and Eq. \eqref{eq:ans2} into the equations of motion for the bulk fields, one can obtain the following static equations
\begin{equation}
	\begin{split}
	\partial_z (f \partial_z u_1)+\frac{1}{r}\partial_r (r\partial _ru_1)&=u_1 \left(-\frac{A_t^2}{f}+\frac{A_\theta^2}{r^2}+\nu\cdot u_2^2+z\right), 
	\\
	\partial_z (f \partial_z u_2)+\frac{1}{r}\partial_r (r\partial _ru_2)&=u_2 \left(-\frac{A_t^2}{f}+\frac{(A_\theta-1)^2}{r^2}+\nu\cdot u_1^2+z\right), 
	\\ 
f \partial_{z}^{2} A_{t}+\partial_{r}^{2} A_{t}+\frac{1}{r} \partial_{r} A_{t}&=2 A_{t} (u_1^2+u_2^2), \\
\partial_{z}\left(f \partial_{z} A_{\theta}\right)+\partial_{r}^{2} A_{\theta}-\frac{1}{r} \partial_{r} A_{\theta}&=2 A_{\theta} (u_1^2+u_2^2).  
\end{split}\label{eq:3}
\end{equation} 
To solve such equations, we are also required to impose some appropriate boundary conditions. Speaking specifically,  we shall impose the following Dirichlet boundary conditions 
\begin{equation}
	A_t \big|_{z=0}=\mu ,~ A_\theta\big|_{z=0}=0,~u_{1,2}\big|_{z=0}=0
\end{equation}
at the AdS boundary. In addition, to eliminate the singular terms in the above  equations of motion, we also impose
\begin{equation}
	A_t\big|_{z=1}=0
\end{equation}
on the black hole horizon. On the other hand, at the spatial boundaries, the behaviors of $u_{1,2}$ are expected to be exchanged with each other, and 
the regularity conditions of the fields near the coordinate singularity $r=0$ also need to be incorporated. As a result, we have the corresponding boundary conditions as follows
\begin{align}
	u_1\to 0, ~\partial_r u_{2}\to 0,~ \partial_r A_t\to 0,~ A_\theta\to 0 \quad & \text{as}\quad r\to \infty, \\
		u_2\to 0, ~\partial_r u_{1}\to 0,~ \partial_r A_t\to 0,~ A_\theta\to 0 \quad &\text{as}\quad r\to 0.
\end{align}

\subsection{Numerical results}
\begin{figure}[ht]
\centering
\includegraphics[width=0.5\textwidth]{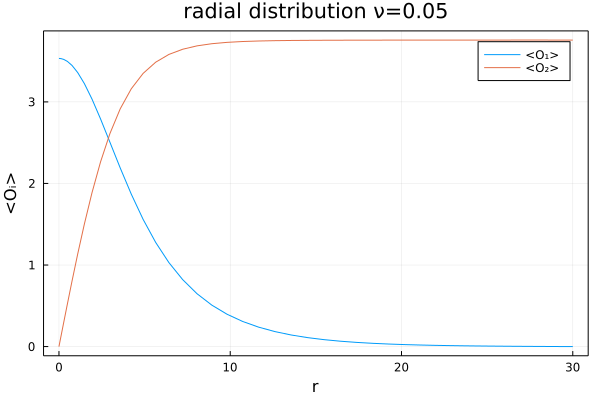}
\caption{The radial profile of the two components of the condensates for the coupling $\nu=0.05$ at $\mu=5.0$.}
\label{fig:1}
\end{figure}
The static solutions can be solved numerically by the pseudo-spectral method. To this end, we make a truncation at $r=R$, where $R$ is chosen to be large enough. For our purpose, we focus on $\mu=5.0$ and set $R=20$. It turns out that the pseudo-spectral discretization with 25 Chebyshev modes in the $z$ direction and 48 Chebyshev modes in the $r$ direction is sufficient for us to obtain the convergent static solutions.

\begin{figure}[ht]
	\centering
		\includegraphics[width=0.4\linewidth]{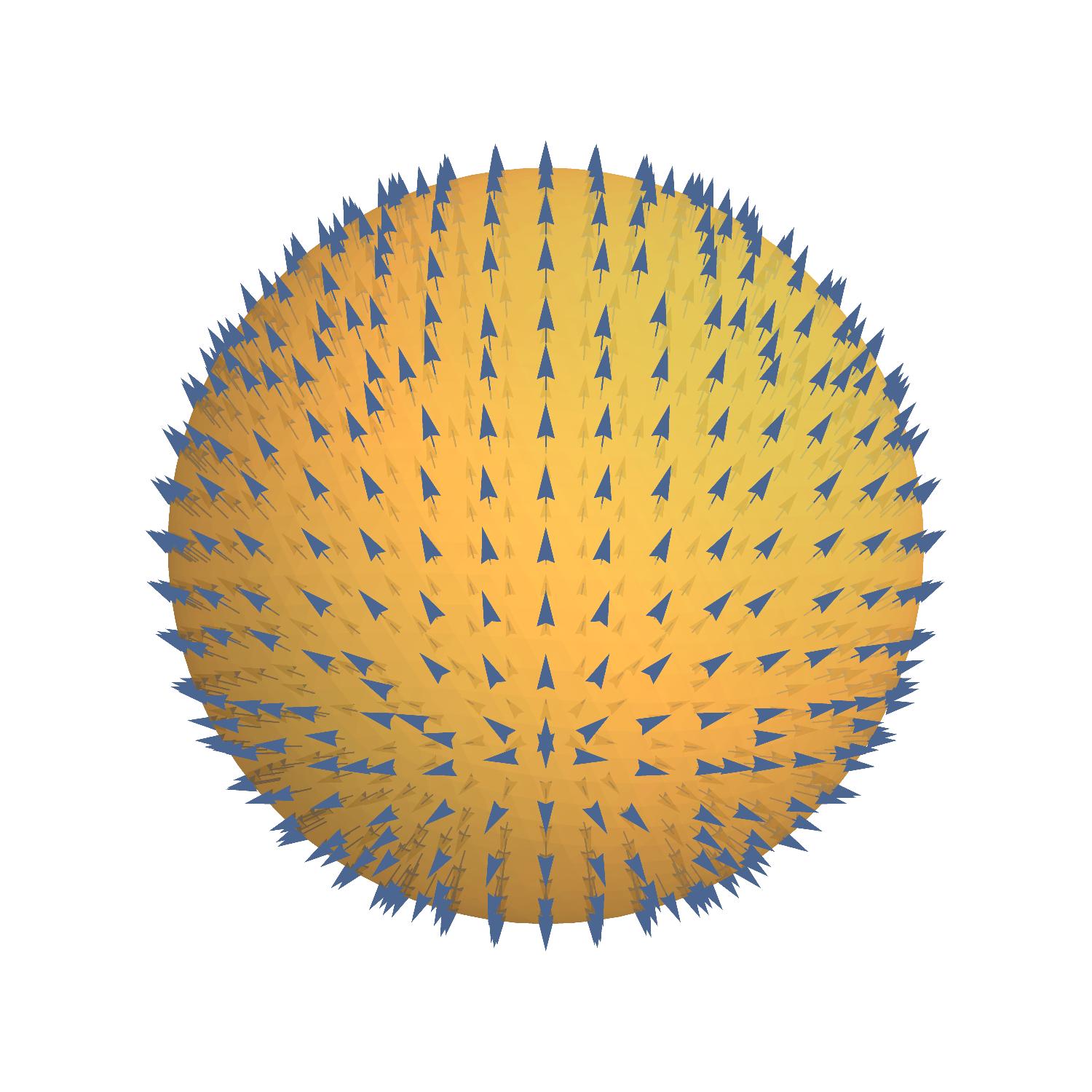}	
	\caption{The hedgehog view of the unit pseudo-spin vector field for $\nu=0.03$ at $\mu=5.0$. }
	\label{fig:2}
\end{figure}
We first plot a typical radial profile of order parameters for the monopole-type baby skymion in Fig.~\ref{fig:1},  
where $\langle\mathcal{O}_i\rangle$ can be obtained by computing $\partial_z u_i$ at $z=0$. As expected, the profile of the second order parameter, which has a winding phase $e^{i\theta}$, behaves like a vortex in the one-order-parameter superfluids, while the condensate of the first order parameter is supported in the vicinity of the vortex center like a Gaussian wave packet.

Our monopole-type baby skyrmion  can also be visualized via a ``hedgehog plot'' as usual. To make it, we first resort to the stereographic projection formula \eqref{eq:4} with $\zeta$ replaced by $z=x+iy$ to map the points on the $x$-$y$ plane to the unit sphere in the 3D Euclidean space, and then attach each site with the unit pseudo-spin vector given by Eq. \eqref{eq:4}. The resulting plot is demonstrated in Fig.~\ref{fig:2}. As we see, the origin of the $x$-$y$ plane is mapped to the north pole of the Riemann sphere with the pseudo-spin pointing north, and the spatial infinity is mapped to the south pole with the pseudo-spin pointing south, while other points on the plane are also mapped to the sphere with the corresponding pseudo-spin pointing to various directions. Through such a hedgehog visualization, one can intuitively see that the skyrmion number  of a monopole-type baby skyrmion is one, because the directions of pseudo-spin vectors distributed on the Riemann sphere cover the target sphere only once.\footnote{Numerical calculation of the integral (\ref{eq:7}) also confirms that the skyrmion number $\mathcal{Q}$ of this monopole-type skyrmion is very close to 1 up to a numerical error of order $10^{-5}$.}

\begin{figure}[bt]
 	\begin{subfigure}{.48\textwidth}
 	\centering
 	\includegraphics[width=0.9\linewidth]{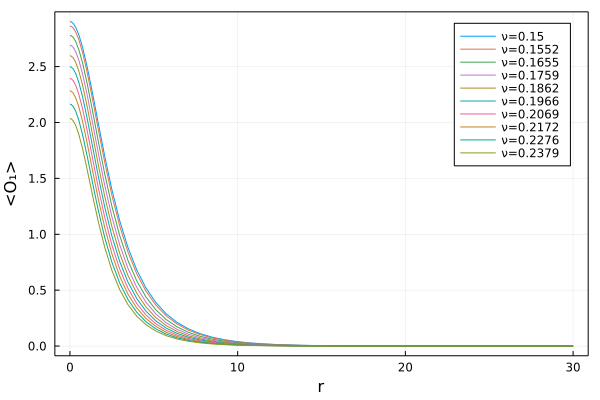}
 	\end{subfigure}
 	\begin{subfigure}{.48\textwidth}
 	\centering
 	\includegraphics[width=0.9\linewidth]{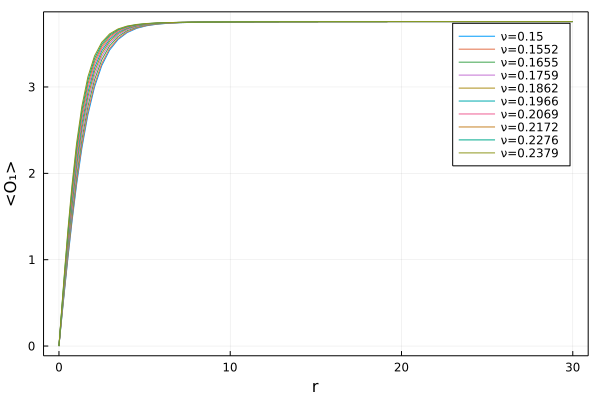}
 	\end{subfigure}
\caption{The radial profile of the two components of the condensates with $\nu$ varied form 0.15 to 0.23. at $\mu=5.0$} \label{fig:vary}
\end{figure}
\begin{figure}[bt]
 \centering
 	\includegraphics[width=0.5\textwidth]{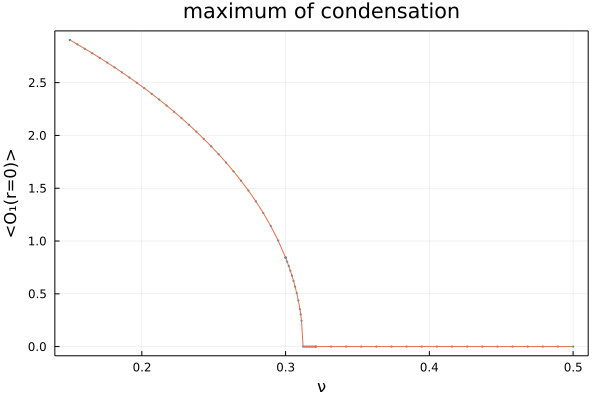}
 	\caption{The maximum of the first component of the condensates decreases to zero as $\nu$ increases to the critical value $\nu_c=0.312$ at $\mu=5.0$, which signals the threshold for the non-existence of the monopole-type baby skyrmion. }
 	\label{fig:crit}
\end{figure}

We have also investigated the effect of the inter-band coupling parameter $\nu$ on the configuration of our monopole-type baby skyrmion. In Fig.~\ref{fig:vary}, we demonstrate the variation of the  radial profile of the two condensates with respect to $\nu$ separately. As one can see, with the increase of $\nu$, the maximal value of  $\langle\mathcal{O}_1\rangle$, located at the origin, is decreased while the second condensate saturates the same maximal value in a steeper manner. When we continue to crank up the coupling parameter $\nu$, the first condensate will be further suppressed and then vanish at some critical coupling $\nu_c\approx 0.312$, which is fixed by plotting the dependence of the maximum of the first condensate on $\nu$ in Fig.~\ref{fig:crit}. This means that the monopole-type baby skyrmion   no longer exists when $\nu\geq\nu_c$. By fitting the curve, we see that near the critical point, there is a square root behavior, that is, 
$\langle\mathcal{O}_1(r=0)\rangle\approx 7.23|\nu-\nu_c|^{1/2}$,
typical of second order phase transitions.

\section{Dipole-type baby skyrmion }\label{section2}

In this section, we are going to study the  $\mathcal{Q}=1$ dipole-type baby skyrmion. The construction of the dipole-type baby skyrmion  by directly solving the static equation turns out to be  rather complicated, for there is no longer a rotational symmetry to exploit. 
As such,  we shall instead resort to the time evolution method to obtain a static configuration at the final stage of the evolution. Actually, the presence of black hole horizon endows the bulk system with a natural dissipation mechanism, which makes the time evolution a natural dynamical relaxation process for obtaining a final static state of interest.

To proceed, below we shall work in the ingoing Eddington-Finkelstein coordinates, i.e.,
\begin{equation}
	d s^{2}=\frac{1}{z^{2}}\left(-f(z) d t^{2}-2 d t d z+d r^{2}+r^{2} d \theta^{2}\right),
\end{equation}
where the full set of equations of motions for our matter sector are presented in the appendix. In what follows, we shall elaborate on our evolution scheme and the initial condition for our numerical simulations before we present the relevant numerical results. 

\subsection{Evolution scheme and initial condition}
In our evolution scheme, the evolution variables are chosen  to be $\{\psi,A_r,A_\theta,\rho\}$, where $\psi$  is given by $\Psi=z\psi$, and
\begin{equation}
	\rho=-\partial_z A_t \big|_{z=0}.
\end{equation}
 Through a simple observation of the evolution equations in the appendix, one can find that the evolution equations \eqref{eq:evol1} to \eqref{eq:evol4} can be cast into the common form as 
 \begin{equation}
     	\partial_t\mathcal{U}=\int_{0}^{z} \mathcal{E}(\mathcal{U},\partial \mathcal{U},A_t)(z') dz'  \quad \text{for}\quad  \mathcal{U}\in\{A_\theta,A_r,\psi\},
 \end{equation} 
 where the boundary conditions 
 \begin{equation}
	\partial_t \mathcal{U} \big|_{z=0}=0 \quad \text{for}\quad \mathcal{U}\in\{\psi,A_\theta,A_r\}
\end{equation}
have been used. On the other hand,  the evolution equation for $\rho$ can be obtained by restricting the mixed evolution equation \eqref{eq:6} to the AdS boundary $z=0$ as
\begin{equation}
		\partial_{t} \rho=-\left.\left(\frac{1}{r} \partial_{z} A_{r}+\partial_{z} \partial_{r} A_{r}+\frac{1}{r^{2}} \partial_{z} \partial_{\theta} A_{\theta}\right)\right|_{z=0}.
\end{equation}
$A_t$ is an intermediate variable, which can be determined by solving the constraint equation \eqref{eq:cons} together with the following boundary conditions
\begin{align}
\partial_z A_t \big|_{z=0}=-\rho, \quad
A_t  \big|_{z=0}=\mu.
\end{align}
Thus the whole evolution system can be formulated as $\partial_t \mathcal{U}=\mathcal{F}(\mathcal{U},A_t(\mathcal{U}))$,  subject to the  boundary conditions at the truncation radius $r=R$ as follows,
\begin{equation}
    \partial_r \psi_{1,2}\big|_{r=R}=0, \quad A_{r,\theta}\big|_{r=R}=0,
\end{equation}
where $\mathcal{U}$ represents the full set of evolution variables.

To proceed,
we still work with $\mu=5.0$, but take $R=15$ for our numerical simulations. The reason why we take $R=15$ rather than $R=20$ as before for the monopole-type baby skyrmion is that the accumulated numerical error, which is monitored by the violation of the constraint equation \eqref{eq:6} in the bulk, will get uncontrolled in the polar coordinates as time goes on if the size of the system is taken too large. With this set-up, we first discretize the spatial directions by the pseudo-spectral method with $30$ Chebyshev modes in the $z$ direction and $85$ Chebyshev modes in the $r$ direction as well as $170$ Fourier modes in the $\theta$ direction, and then the evolution equation can be integrated out by the $4$th order Runge-Kutta method with a typical time step $\Delta t=0.002$, where the code is written in Julia\cite{julia}.

 To set off the above evolution scheme, we need to prescribe an appropriate set of initial conditions. For our purpose,  we like to start with the initial configuration homotopic to the dipole-type baby skyrmion. As such, the initial value for $\psi$ is chosen as follows
\begin{equation}
	\psi(t=0,z,r,\theta)=\binom{\psi_0(z) f(r_1/r_2)e^{i\theta_1}}{\psi_0(z) f(r_2/r_1)e^{i\theta_2}}, \quad  \theta_i=\arctan{\frac{y-y_i}{x-x_i}},\quad r_i=|\mathbf{r}-\mathbf{r}_i|,
\end{equation}
where $\psi_0(z)$ denotes the corresponding static configuration for the homogeneous isotropic superfluid, and  $f(\cdot)=\tanh(\frac{\cdot}{\xi})$ is a core function of a single vortex with $\xi$ the typical size of a single vortex. Similarly, the initial value for $\rho$ is also chosen as that for the homogeneous isotropic superfluid. Finally, the initial data for $A_r$ and $A_\theta$ are set to be zero for simplicity.

\subsection{Numerical results}

 Our numerical simulations indicate that the final state will not settle down to the dipole-type baby skyrmion   when $\nu>0$. 

\begin{figure}[ht]
\centering
\begin{subfigure}{0.3\textwidth}
	\centering
	\includegraphics[width=\linewidth]{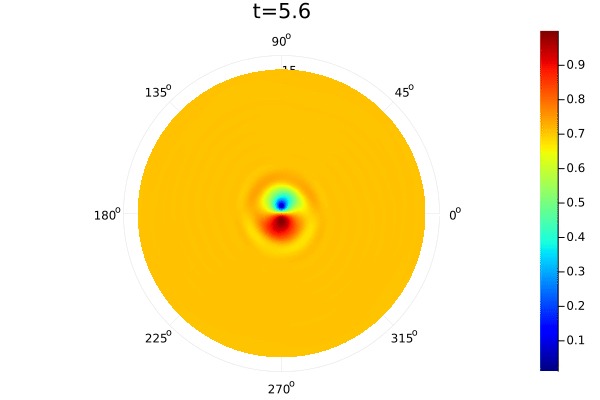}
\end{subfigure}
\begin{subfigure}{0.3\textwidth}
	\centering
	\includegraphics[width=\linewidth]{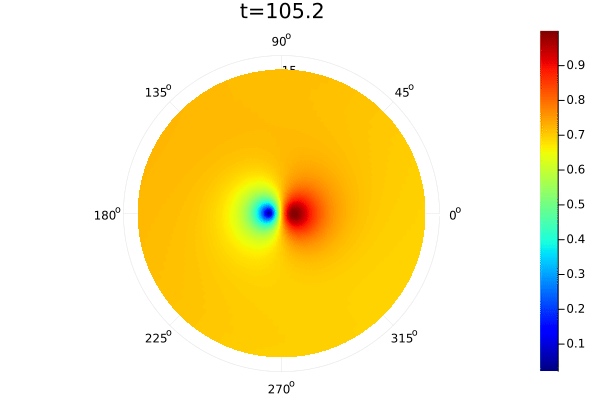}
\end{subfigure}
\begin{subfigure}{0.3\textwidth}
	\centering
	\includegraphics[width=\linewidth]{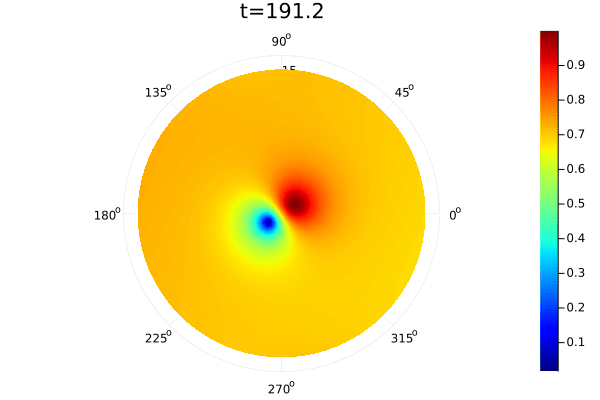}
\end{subfigure}

\centering
\begin{subfigure}{0.3\textwidth}
	\centering
	\includegraphics[width=\linewidth]{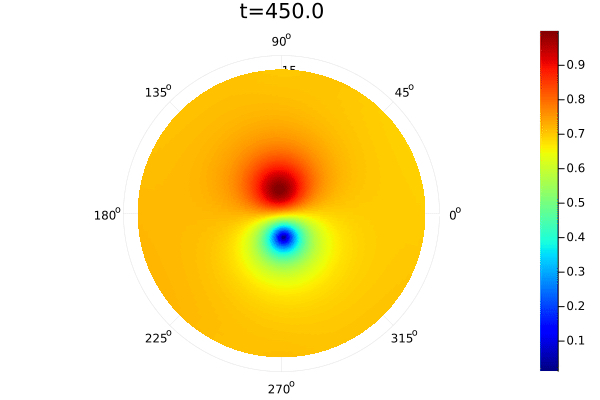}
\end{subfigure}
\begin{subfigure}{0.3\textwidth}
	\centering
	\includegraphics[width=\linewidth]{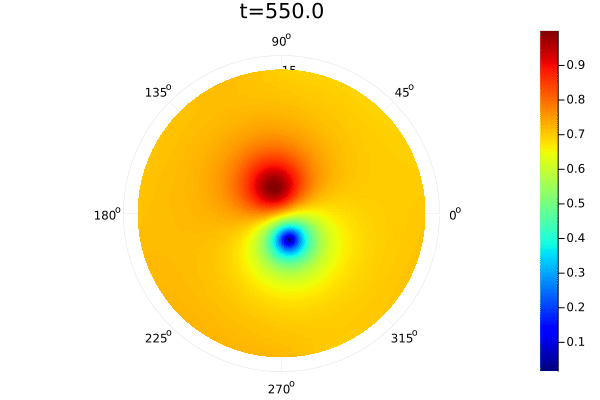}
\end{subfigure}
\begin{subfigure}{0.3\textwidth}
	\centering
	\includegraphics[width=\linewidth]{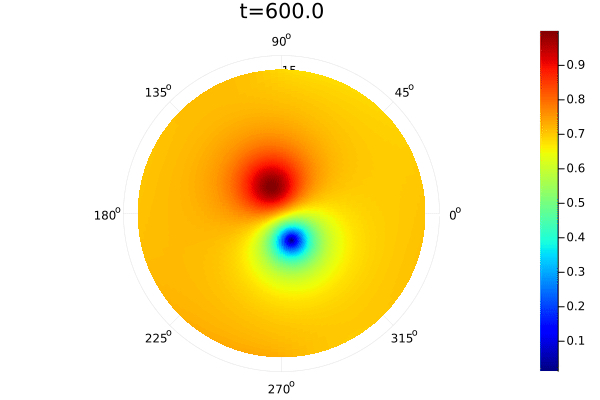}
\end{subfigure}
\caption{Snapshots of the temporal evolution of the condensate for $\nu=-0.15$ at $\mu=5.0$. }
\label{fig:5}
\end{figure}

Here we demonstrate the typical temporal evolution of our system in Fig.~\ref{fig:5} with $\nu=-0.15$, where we plot the distribution of the condensate with the color indicating the normalized value of the first condensate i.e., $\langle\mathcal{O}_1\rangle/\sqrt{\langle\mathcal{O}_1\rangle^2+\langle\mathcal{O}_2\rangle^2}$, such that red color represents $1$ while the blue color stands for $0$. As one can see, in each snapshot, there are a blue core and a red core. The center of the blue core is exactly where the vortex of $\langle\mathcal{O}_1\rangle$ resides, which corresponds to the south pole of the target sphere. Similarly, the center of the red core, which is the vortex center of $\langle\mathcal{O}_2\rangle$, corresponds to the north pole.

At the initial stage, the two poles of the dipole baby skyrmion are set very close to each other, with $\mathbf{r}_1=-\mathbf{r}_2$ for the sake of symmetry. At early times they are seen to begin repelling each other as well as rotating anti-clockwise around each other. As time goes on, such a process slows down gradually. After $t=500$, the system settles down to a final static state, which can be regarded as the static baby skyrmion of the dipole type we desire. This result also indicates the dynamical stability of the resulting dipole-type baby skyrmion.

\begin{figure}[ht]
 \centering
 	\includegraphics[width=0.5\textwidth]{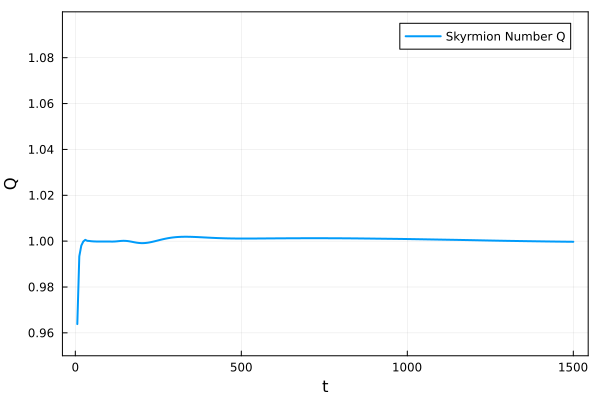}
 	\caption{Skyrmion number $\mathcal{Q}$ as a function of time. In late times, $\mathcal{Q}$ converges to 1 as it should be the case.}
 	\label{fig:nqs}
\end{figure}

As depicted in Fig.~\ref{fig:nqs}, we also monitor the skyrmion number $\mathcal{Q}$ at every time step of evolution. As one can see, the numerically calculated topological charge $\mathcal{Q}$ is indeed very close to 1 with a negligible error of order $10^{-4}$ at late times\footnote{The initial deviation from $1$ comes essentially from the non-smoothness of the initial data, which  dies out by the black hole at late times.}.

It is noteworthy that we have never seen our temporal evolution for $\nu<0$ end up with the monopole-type baby skyrmion as the final state, which indicates the non-existence of the monopole-type baby skyrmion in this region\footnote{This also echos the fact we fail to construct the monopole type of baby skyrmion in the $\nu<0$ region by solving the static equations of motion.}.

\section{Discussion}

In this paper, we have succeeded in constructing both the monopole-type  baby skyrmion  and the dipole-type one in the holographic two-component superfluid model by directly solving the static equation of motion and resorting to the temporal evolution scheme, respectively. In addition, our numerical result indicates that the existence of such baby skyrmions depends on the  interaction between the two components. In particular, the monopole-type baby skyrmion  can exist only in the coupling parameter $0<\nu<\nu_c$ and the dipole-type baby skyrmion  can exist only in $\nu<0$.  Such a result is reminiscent of that obtained for the baby skyrmion model, where the existence of such baby skyrmions also depends on the specific interaction term\cite{Leask:2021hzm}. Hence, it would be highly rewarding to develop an effective boundary description of our bottom-up bulk model to deepen our understanding of such a similar behavior. 

We conclude our paper with a few additional issues worthy of further investigation. The first direct generalizations of our work include examining how our findings are affected by choosing different mass parameters $m_1$ and $m_2$ of our scalar doublet, giving up on the assumption that the gauge couplings of $\Psi_1$ and $\Psi_2$ with the $U(1)$ gauge field are the same, or taking into account the backreaction of the matter fields onto the background geometry. Another direction is to extend our work to the three spatial dimensions, where  skyrmion corresponds to the local field configuration  which induces a topologically nontrivial mapping from $\mathbb{R}^3$ to $S^3$\cite{skyrme1962unified,skyrme1994selected,kotiuga2021continuum}. There exist some attempts in the context of 3D two-component GL theory\cite{Battye:2001ec,Kawakami:2012zw}, but we expect to report our result in the holographic model in the near future. Last but not least, there have been a bunch of endeavors devoted to the statistical law of the quantum turbulence of vortices such as Kolmogorov scaling law\cite{Adams:2012pj,Lan:2016},
our current work opens a novel avenue for one to investigate the statistical law of the quantum turbulence associated with these baby skyrmions by holography, where some new features may be found.

\acknowledgments
We are grateful to Yongliang Ma and Sven Bjarke Gudnason for helpful discussions. This work is partially supported by the National Key Research and Development Program of China Grant No. 2021YFC2203001 as well as the Natural
Science Foundation of China under Grants No.~11975235, No.~12035016, and No.~12075026.

\appendix

\section{Appendix: Equations of motion in the ingoing Eddington-Finkelstein coordinates} \label{sec:equations}
With the scalar doublet $\Psi=z(\psi_1,\psi_2)$ and the axial gauge $A_z=0$, the whole set of equations of motion in the ingoing-Eddington-Finkelstein coordinates can be classified as the following evolution equations

\begin{align}
\partial_z\partial_t \psi_1&=(-\Phi+\nu|\psi_2|^2)\psi_1 /2+(1/r^2-2iA_r)\partial_r \psi_1/2+ \\
&\partial_r^2\psi_1/2-i A_\theta \partial_\theta \psi_1/r^2+\partial_\theta^2\psi_1/r^2+(i A_t+F_{r\theta}) \partial_z \psi_1+ F_{r\theta} ~\partial_z^2 \psi_1/2,
 \label{eq:evol1} \\ 
\partial_z\partial_t \psi_2&=(-\Phi+\nu|\psi_1|^2)\psi_2 /2+(1/r^2-2iA_r)\partial_r \psi_2/2+ \\
&\partial_r^2\psi_2/2-i A_\theta \partial_\theta \psi_2/r^2+\partial_\theta^2\psi_2/r^2+(i A_t+F_{r\theta}) \partial_z \psi_2+ F_{r\theta} ~\partial_z^2 \psi_2/2,
\label{eq:evol2} \\
 \partial_z\partial_t A_r&=-A_r (\psi^\dagger\psi)+\text{Im}(\psi^\dagger\partial_r\psi)-
 \partial_\theta F_{r\theta}/2r^2+
 \partial_z(f\partial_z A_r+\partial_r A_t)/2,
 \label{eq:evol3}  \\
 \partial_z\partial_t A_\theta&=-A_\theta (\psi^\dagger\psi)+\text{Im}(\psi^\dagger\partial_\theta\psi)-F_{r\theta}/2r^2+\partial_r F_{r\theta}/2+
 \partial_z(f\partial_z A_\theta+\partial_\theta A_t)/2,
 \label{eq:evol4}
\end{align}
where 
\begin{equation}
	\Phi=A_r^2+A_\theta^2/r^2+z-i\partial_z A_t+i(A_r/r+\partial_r A_r+\partial_\theta A_\theta/r^2),\quad
	F_{r\theta}=\partial_r A_\theta-\partial_\theta A_r,
\end{equation}
as well as 
one constrain equation 
\begin{equation}
\partial_z^2 A_t=-2\text{Im}(\psi^\dagger\partial_z\psi)+\partial_z (A_r/r+\partial_r A_r+\partial_\theta A_\theta/r^2) \label{eq:cons}
\end{equation} 
and one mixed evolution equation
\begin{equation}
\begin{gathered}
\partial_{t} \partial_{z} A_{t}=-\frac{1}{r} \partial_{t} A_{r}-\partial_{t} \partial_{r} A_{r}-\frac{1}{r^{2}} \partial_{t} \partial_{\theta} A_{\theta}+\frac{f}{r} \partial_{z} A_{r}+f \partial_{z} \partial_{r} A_{r}+\frac{f}{r^{2}} \partial_{z} \partial_{\theta} A_{\theta}+\partial_{r}^{2} A_{t} \\
+\frac{1}{r} \partial_{r} A_{t}+\frac{1}{r^{2}} \partial_{\theta}^{2} A_{t}+\text{Im}(\psi^{\dagger} \partial_{t} \psi)-2 A_{t} \psi^\dagger \Psi-2f\text{Im}(\psi^\dagger\partial_{z} \psi).
\end{gathered}\label{eq:6}
\end{equation}

\bibliographystyle{JHEP}
\bibliography{skbb}
\end{document}